\documentstyle[12pt,epsf]{article}
\def\be{\begin{equation}}
\def\ee{\end{equation}}
\def\bea{\begin{eqnarray}}
\def\eea{\end{eqnarray}}

\def\br{\begin{thebibliography}{abc}}
\def\er{\end{thebibliography}}

\def\bar#1{\overline{#1}}

\def\a{\alpha}
\def\b{\beta}
\def\c{\raisebox{.4ex}{$\chi$}}

\def\e{\epsilon}

\newcommand{\ZZ}{ Z \!\!\! Z}
\newcommand{\comm}{{\rm l}\!\!\!{\rm C}}
\def\real{\relax{\rm I\kern-.18em R}}
\def\com{\relax\,\hbox{$\inbar\kern-.3em{\rm C}$}}

\def\br{{\bf R}}

\def\inbar{\,\vrule height1.5ex width.4pt depth0pt}
\def\IC{\relax\,\hbox{$\inbar\kern-.3em{\rm C}$}}
\def\ID{\relax{\rm I\kern-.18em D}}
\def\IF{\relax{\rm I\kern-.18em F}}
\def\IH{\relax{\rm I\kern-.18em H}}
\def\II{\relax{\rm I\kern-.17em I}}
\def\I1{\relax{\rm 1\kern-.28em l}}
\def\IN{\relax{\rm I\kern-.18em N}}
\def\IP{\relax{\rm I\kern-.18em P}}
\def\IQ{\relax\,\hbox{$\inbar\kern-.3em{\rm Q}$}}
\def\IZ{\relax\,\hbox{$\inbar\kern-.3em{\rm Z}$}}
\def\R{\relax{\rm I\kern-.18em R}}
\font\cmss=cmss10 \font\cmsss=cmss10 at 7pt
\def\Z{\relax\ifmmode\mathchoice
{\hbox{\cmss Z\kern-.4em Z}}{\hbox{\cmss Z\kern-.4em Z}}
{\lower.9pt\hbox{\cmsss Z\kern-.4em Z}}
{\lower1.2pt\hbox{\cmsss Z\kern-.4em Z}}\else{\cmss Z\kern-.4em
Z}\fi}
\def\ca{{\cal A}}
\def\cb{{\cal B}}
\def\cc{{\cal C}}

\def\ch{{\cal H}}

\def\cn{{\cal N}}
\def\co{{\cal O}}

\def\bar#1{\overline{#1}}

\def\Hat#1{\rlap{\kern.10em$\widehat{\phantom G}$}#1}
\def\HAt#1{\rlap{\kern.05em$\widehat{\phantom G}$}#1}

\def\czp#1{\rlap{\kern.1em$\widehat{\phantom{G\vrule height.8em}}$}#1{}}
\def\Czp#1{\rlap{\kern.05em$\widehat{\phantom{G\vrule height.8em}}$}#1{}}

\newcommand{\sect}[1]{\setcounter{equation}{0}\section{#1}}

\def\sxn#1{\bigskip\medskip \sect{#1} \smallskip
                                                 }

\begin{document}

\thispagestyle{empty}
\setcounter{page}{0}


\begin{flushright}
SU-4240-716\\
February 2000
\end{flushright}

\vglue 0.6cm

\centerline {{\Large{\bf Classical Topology and Quantum
States}}\footnote{Talk
presented at the Winter Institute on Foundations of Quantum
Theory and Quantum Optics,S.N.Bose National Centre for Basic 
Sciences,Calcutta,January
1-13,2000.}}

\vglue 0.6cm
                        
\centerline {\large  A.P. Balachandran}
\vglue 0.5cm
{\centerline{\it {Department of Physics, Syracuse University,}}
\centerline{\it {Syracuse, NY 13244-1130, USA}}

\vglue 1cm

\centerline {\bf Abstract}
\vglue 0.6cm

     Any two infinite-dimensional (separable) Hilbert spaces are
unitarily
isomorphic.  The sets of all their self-adjoint operators are also
therefore unitarily
equivalent.  Thus if all self-adjoint operators can be observed, and if
there is no further major axiom in quantum physics than those formulated
for example in Dirac's `Quantum Mechanics', then a quantum physicist
would not be able to tell a  torus from a hole in the ground.  We argue
that there are indeed such axioms involving observables with smooth time
evolution:  they contain commutative subalgebras
from which the spatial slice of spacetime with its
topology (and with further refinements of the axiom, its $C^K-$ and
$C^\infty-$
structures) can be reconstructed using Gel'fand - Naimark theory and its 
extensions.
Classical topology is an attribute of only certain quantum observables
for
these axioms, the spatial slice emergent from quantum physics
getting progressively less differentiable with increasingly higher
excitations of energy and eventually altogether ceasing to exist.  After
formulating these axioms, we apply them to show the possibility of
topology change and
to discuss quantized fuzzy topologies.  Fundamental issues concerning the
role of time in quantum physics are also addressed.

\newpage

\vglue 0.6cm                     
\sxn{{\bf Introduction}}
\vglue 0.5cm
	
Conventional expositions of classical physics assume that the concept of
the spatial slice $Q$  and its topological and differential geometric
attributes
are somehow known , and formulate dynamics of particles or fields using   
$Q$ and further metaprinciples like locality and causality. The space
$Q$ thus
becomes an irreducible background , immune to analysis, for a classical
physicist, even though it is an indispensable ingredient in the
formulation of physical theory.

Quantum physics is a better approximation to reality than is classical
physics.
Still, models of quantum physics are seldom  autonomous, but are rather
emergent from a classical substructure. Thus we generally formulate a
quantum model by canonical or path integral quantisation of a classical
Lagrangian based on the space $Q$ . We thus see that $Q$ and
its
properties are tamely accepted , and they are not subjected to physical
or mathematical analysis, in such conservative quantum physics too.

Classical topology is in this manner incorporated in conventional
quantum physics by formulating it using 
smooth functions on $Q$. There is reason to be uneasy with this method 
of encoding classical data in quantum physics.
In quantum theory, the
fundamental
physical structure is the algebra of observables, and it would be greatly
more
satisfactory if we can learn if and how operator algebras describe
classical
topology and its differential attributes.

	This note will report on certain ongoing research with several
colleagues concerning this question which is fundamentally an enquiry
into the nature of space and time in quantum physics.
Some of our ideas have already been published elsewhere
\cite{puri,bbellst,bbms,bc}. Our
work touches both on issues of relevance to quantum gravity such as the
meaning of ``quantized topology'' and the possibility of topology change,
and on topics of significance for foundations of quantum physics.  I
think that we have  progressively approached a measure of precision in
the formulation of relatively inarticulated questions, but our responses
are still  tentative and lacking in physical and mathematical
completeness and rigor.

\vglue 0.6cm
\sxn{ {\bf The Problem as a Parable}}
\vglue 0.5cm

We restate the problem to be addressed here.It is best introduced as a little 
story
about a quantum baby.  The story will set the framework for the rest of
the talk.  Its proper enjoyment calls for a willing suspension of
disbelief for the moment.

All babies are naturally quantum, so my adjective for the baby can be
objected to as redundant and provocative, but it invites attention to a
nature of infants of central interest to us, so let us leave it there.

\vglue 0.6cm
\noindent {\bf Parable of the Quantum Baby}
\vglue 0.5cm

Entertain the conjecture of a time, long long ago, when there lived a
quantum baby of cheerful semblance and sweet majesty.  It was brought up
by its doting parents on a nourishing diet of self-adjoint operators on a
Hilbert space.  All it could experience as it grew up were their mean
values in quantum states.  It  did not have a clue when it was little
that there is our classical world with its topology, dimension and
metric.  It could not then tell a torus from a hole in the  ground.

Yet the baby learned all that as it grew up.  

And the wise philosopher is
struck with wonder:  How did the baby manage this amazing task?  
 
For the problem is this: Even in a quantum theory emergent  from a smooth
classical configuration space $Q$, there is no need for a wave
function $\psi$, or a probability density $\psi^*\psi$, to be continuous
on
$Q$.  It is enough that the integral $\int \omega \psi^*\psi$ over
$Q$ for an appropriate volume form $\omega$ is finite.
Probability
interpretation  requires no more.

But if the baby can observe all self-adjoint operators with equal ease,
and thereby prepare all sorts of discontinuous quantum states, how then
does it ever learn of $Q$, its topology and its differential
attributes ? The problem is even worse: We shall see below that any two
(separable)Hilbert spaces are isometric so that there is only one abstract
Hilbert space.

This then is our central question.  All that follows is charged with its
emotional content, and comes from trying to find its answer.

\vglue 0.6cm
\sxn{{\bf  Another Statement}}
\vglue 0.5cm

We can explain the baby problem in yet another way.

In quantum physics,observables come from bounded operators
on a (separable \cite{rs}) Hilbert space $\ch$.  [We will deal only
with
separable Hilbert spaces.]  The latter is generally infinite-dimensional.

But all infinite-dimensional Hilbert spaces are
isomorphic, in fact unitary so.  If $|n>^{(i)}(n\in {\IN})$ gives the
orthonormal basis for the Hilbert space $\ch^{(i)} (i=1,2)$, we can
achieve this equivalence by setting $|n>^{(2)}=V|n>^{(1)}$.  That being
so, any operator $A^{(1)}$ on $\ch^{(1)}$ has a corresponding operator 
$A^{(2)}= VA^{(1)}V^{-1}$ on $\ch^{2)}$. 

How then does a quantum baby tell a torus from a hole in the ground?

{\em Without further structure in quantum physics besides those to
be found in standard text books, this task is in fact entirely beyond the
baby.}

In conventional quantum physics of particles say,we generally start from
smooth functions
(or smooth sections of hermitean vector bundles) on $Q$ and
complete them into a Hilbert space $\ch$ using a suitable scalar product.
In this way, we somehow incorporate knowledge about $Q$ right at
the start.

But this approach requires realizing $\ch$ in a particular way, as square
integrable functions (or sections of hermitean vector bundles)on $Q$.  The 
presentation of $\ch$ in this
manner is reminiscent of the presentation of a manifold in a preferred
manner, as for instance using a particular coordinate chart.

Can we give a reconstruction of $Q$ in an intrinsic way?  What new
structures are needed for  this purpose?

In the scheme we develop as a response to these questions, $Q$
emerges with its $C^{\infty}-$ structure only from certain observables,
 {\em topology and differential features  being attributes of
particular classes of observables and not universal properties of all
observables.}
Thus $Q$ emerges as a manifold only if the high energy components
in the observables are suppressed. When higher and higher
energies are excited, it gets more and more rough and eventually altogether 
ceases
to exist as a topological space modelled on a manifold.  Here by becoming
more rough we mean that $C^\infty$ becomes $C^K$
and correspondingly the $C^\infty-$ manifold $Q$ becomes a
$C^K-$ space $Q^K$.

The epistemological problems we raise here are not uniquely quantal.They
are encountered in classical physics too , but we will not discuss them
here.

\vglue 0.6cm
\sxn{ {\bf  What is Our Quantum System?}}
\vglue 0.5cm

The system we consider is generic. If $K$ is the configuration space of a
generic system,such as that of a single particle or a quantum field,its
algebra of observables normally contains the algebra $C^{\infty}(Q)$ of
smooth functions on the spatial slice $Q$.For a charged field, for
example,suitably smeared charge, energy and momentum densities can
generate this algebra. That is ( provisionally) enough  for our central
goal of recovering $Q$ from quantum observables.

\vglue 0.6cm
\sxn{{\bf  Time is Special}}
\vglue 0.5cm

We have to assume that time evolution is given as a  unitary
operator $U(t)$ which is continuous in $t$.  Our analysis needs this
input. Time therefore persists
as an {\em a priori} irreducible notion even in our quantum approach.  It
would
be very desirable to overcome this limitation. (See \cite{CoRo} in this 
connection.)

There is more to be said on time, its role in measurement theory and as
the mediation between quantum and classical physics.  There are
brief remarks on these matters below.

It is true that in so far as our main text is concerned, $U(t)$ or the
Hamiltonian can be substituted by spatial translations, momenta or other
favorite observables.  But we think that time evolution is something
special, being of universal and  central interest to science.  It is for
this reason that we have singled out $U(t)$.

\vglue 0.6cm
\sxn{ {\bf  The Gel'fand-Naimark Theory}}
\vglue 0.5cm

The principal mathematical tool of our analysis involves this remarkable
theory \cite{fd} and, to some extent its developments in Noncommutative
Geometry \cite{co,col1,con,Landi,ma}.  We shall now give a crude and short
sketch of this
theory.

A $C^*$-algebra $\ca$ with elements $c$ has the following
properties: a) It is an algebra over $\comm$. b) It is 
closed under an antinvolution $*$:

\be
*:c_j \in \ca\Rightarrow c^{*}_{j} \in \ca, ~c^{**}_{j} =
c_j,~
(c_1c_2)^*=c^{*}_{2}c^*_{1},~(\xi c_j)^*=\xi^*c^{*}_{j},
\ee

\noindent where $\xi$ is a complex number and $\xi^*$ is its complex
conjugate.
c) It has a norm $|| . || $ with the properties 
 $|| c^*|| = || c ||,~ || c^* c|| = || c ||^2$
for $ c \in \ca$.
d) It is complete under this norm.   

A $*-$ representation $\rho$ of $\ca$ on a Hilbert space $\ch$ is 
a representation of $\ca$ by a $C^*$-algebra of bounded operators on
 a  Hilbert space \cite{rs2} with the following features:
i)  The $*$ and norm for $\rho (\ca)$ are the operator adjoint
$\dagger$
and
operator norm (also denoted by $||\cdot ||$). ii)
$\rho(c^*)=\rho(c)^\dagger$.

$\rho$ is said to be a $*$-homomorphism because of ii).  We can also in a
similar manner speak of $*$-isomorphisms.  

We will generally encounter
$\ca$ concretely as  an algebra of operators.  In any case, we will
usually omit the symbol $\rho$.

Note that a $*$-algebra (even if it is not $C^*$) is by definition closed
under an antinvolution $*$.

Let ${\cc} $ denote a commutative  $C^*$-algebra. 
Let $\{ x \}$ denote its space of inequivalent irreducible
$*$-representations
(IRR's) or its spectrum. [So $a\in \cc \Rightarrow x(a) \in
\comm $.]
The Gel'fand-Naimark
theory then makes the following striking assertions: $\a)$ There is
a natural topology on $\{x\}$ making it into a Hausdorff topological
space
\cite{ency} $Q^0$. [We will denote the IRR's prior to introducing topology
by
$\{x\}$ and after doing so by $Q^\cdot$ with suitable superscripts.$Q$ is 
the same as $Q^{\infty}$ below.] 
$\b$) Let $\cc^0(Q)$ be the $C^*$-algebra of~ $\comm$-valued
continuous
functions on $Q$.  Its $*$ is complex conjugation and its norm
$||\cdot ||$ is the supremum norm, $|| \phi ||= \sup_{x\in Q^{0}} |\phi
(x)|.$
Then $\cc^0(Q)$ is $*$-isomorphic to $\cc$.

We can thus identify $\cc^0(Q)$ with $\cc$ , as we will often do.

The above results can be understood as follows.  By ``duality'', the
collection of $x(a)$'s for all $x$ defines a function $a_c$ on $\{ x\}$
by $a_c(x):=x(a)$. $ a_c$ is said to be the Gel'fand transform of $a$.

$\{x\}$ is
as
yet just a collection of points with no
topology.  How can we give it a natural topology? We want $a_c$ to be
$C^0$ in this topology.  Now the set of zeros of a continuous function is
closed.  So let us identify the set of zeros $C_a$ of each $a_c$ with a
closed set:

\be
C_a = \{x:x(a)\equiv a_c(x)=0\}.
\ee

\noindent The topology we seek is given by these closed sets.  The
Gel'fand-Naimark
theorem then asserts $\a$) and $\b$) for this topology, the isomorphism 
$\cc \rightarrow\cc^0(Q) $ being $a \rightarrow a_c$.

A Hausdorff topological space can therefore be equally well described by
a commutative  $C^*$ -algebra $\ {\cc}$, presented for example using
generators.  That would be an intrinsic coordinate-free description of
the space and an alternative to using coordinate charts.

A $C^K$ - structure can now be specified by identifying an appropriate
subalgebra $\cc^K$ of $\cc \equiv \cc^0$ and declaring that the
$C^K$ - structure is the one for which $\cc^K$ consists of $K$-times
differentiable functions.  [$\cc^K$ is a $*-$, but not a $C^*-$,
algebra for $K>0$, as it is not complete..] The corresponding $C^K$-space 
is $Q^K$.
For $K=\infty$, we get the manifold $Q^\infty$.  We have the inclusions

\be
\cc^\infty \subset ... \subset \cc^K ... \subset \cc^0 \equiv \cc
\ee

\noindent where

\be
\bar {\cc}^{(\infty)} = \bar{\cc}^{(K)} = \bar {\cc}\equiv\cc,
\ee

\noindent the bar as usual denoting closure.In contrast, $Q^\infty$ and 
$Q^K$ are all the same as sets,
being
$\{x\}$.

A dense $*$-subalgebra of a $C^*$-algebra $\cc$ will be denoted
by
$\cc^{\cdot}$, the superscript highlighting some additional
property.  The algebras $\cc^K$ are  such examples.

{\bf Example 1}: Consider the algebra $\cc$ generated by the
identity, an
element $u$ and its inverse $u^{- 1}$. Its elements are
$a= \sum_{N \in {\ZZ}} { \alpha_{N}u^{N}}$
where $\a_N$'s are complex numbers vanishing rapidly in $N$ at $\infty$.
The $*$
is defined by $u^*=u^{-1}, a^*=\sum a^{*}_{N} u^{-N}$.  As $\cc$ has
identity $\I1$ , there is a natural way to define inverse $a^{-1}$ too :
$a^{-1}$ is that element of $\cc$ such that $a^{-1}a=aa^{-1} = {\I1}$.
There
is also a canonical norm $||.||$ compatible with properties c)
\cite{co,con}:
$||a||$ = Maximum of $|\lambda|$ such that $a^*a-|\lambda|^2$ has no
inverse.

The space $Q$ for this $\cc$ is just the circle $S^1$, $u_c$
being the function with value $e^{i\theta}$ at $e^{i\theta} \in S^1$.

If similarly we consider the algebra associated with $N$ commuting
unitary
elements, we get the $N$-torus $T^N$.  If  for $N=2$, the generating
unitary elements do not commute, but fulfill  $u_1u_2= \omega u_2 u_1,
\omega$ being any phase, we get the noncommutative torus
\cite{cor,co}. It is the ``rational'' or ``fuzzy'' torus if $\omega^K=1$
for
some $K \in
{\ZZ}$, otherwise it is ``irrational'' \cite{ma,seiberg}.

\vglue 0.6cm
\sxn{{\bf States and Observables}}
\vglue 0.5cm

The formulation of quantum physics best suited for the current discussion
is based on the algebra $\cb$ of bounded observables and states $\omega$ 
on $\cb$.
$\cb$ has a *-operation ( anti-involution) and $\omega (b)\in \comm $ for 
$b\in \cb$ with $\omega (b^{*}b)$ $\geq 0 $, $\omega (\I1)=1$. 
$\omega $ can be thought of as the density matrix 
describing the ensemble and $b$ the operator whose mean value is being 
measured. The Gel'fand-Naimark-Segal (GNS) construction lets us recover the 
Hilbert space formulation from $\omega$ and $b$.

\vglue 0.6cm
\sxn{{\bf  Instantaneous Measurements and Classical Topology}}
\vglue 0.5cm

Time in $\em{conventional}$ quantum physics has a unique role.It is not a 
quantum variable, and all elementary quantum observations are instantaneous.

Now elementary measurements-those instantaneous in time- can only measure 
commuting observables.Thus the probabilty of finding the value $a$ for the
observable $A$ at time $t-\e$ and then $b$ for $B$ at $t+\e$ is 
$\omega (P_a(t-\e)P_b(t+\e)P_a(t+\e))$,where $P_{a,b}$ are projectors at 
indicated times. If the order is reversed, the answer is 
$\omega (P_a(t-\e)P_b(t+\e)P_a(t+\e))$ . They do not coincide as 
$\e\rightarrow0$ unless  $P_{a}(t)P_{b}(t)=P_{b}(t)P_{a}(t)$, that is 
$AB=BA$.As experiments cannot resolve time sequence if $\e$ is small enough,
we cannot  consistently assign joint probabilities to noncommuting observables
in elementary measurements.

Thus from instantaneous measurements, we can extract commutative
 $C^*$-algebras and therefrom Hausdorff topological spaces.

{\em If ``commutation'' is classical, then instantaneous measurements and 
Hausdorff spaces ( the stuff of manifolds ) are also partners in this 
classicality.}

It is known that a state $\omega$ restricted to a commutative $C^*$-algebra is
equivalent to a classical probabilty measure on its underlying topological
space.
As a wave function $|\psi>$ thus is equivalent to a classical probability 
measure for an
instantaneous measurement (which any way is the only sort of measurement
discussed in usual quantum physics), there is no need to invoke 
``collapse of wave packets'' or similar hypotheses for its interpretation.
The uniqueness of quantum measurement theory then consists in the special
relations it predicts between outcomes of measurements of different
commutative  algebras $\cc_1$ and $\cc_2$.  These relations are often
universal,
being independent of the state vector $|\psi>$.

Such a point of view of quantum physics, or at least a view close to it,
has been advocated especially by Sorkin \cite{sor4}.

Thus we see that instantaneous measurements are linked both to classical
topology and to classical measurement theory.

But surely the notion of {\em instantaneous measurements} can only be an
idealization.  Measurements must be extended in time too, just as they are
extended in space.  But we know of no fully articulated theory of
measurements extended in time, and maintaining quantum coherence during
its
duration, although interesting research about these matters exists \cite{ha1}.

A quantum theory of measurements extended in time, with testable
predictions, could be of fundamental importance. We can anticipate that it
will involve 
noncommutative algebras $\cn$ instead of
commutative algebras, the hermitean form $\psi^\dagger\chi$ for 
the appropriate vectors
$\psi,
\chi$  in the Hilbert space being valued in $\cn$.  
Such quantum theories were
encountered in \cite{bbellst}.  Mathematical tools for their further 
development are
probably available in Noncommutative Geometry \cite{co,col1,con,Landi,ma}.

But we are hardly done , we do not have $Q$ as a manifold , or its dimension
etc.

\vglue 0.6cm
\sxn{{\bf  What Time Evolution Tells Us}}
\vglue 0.5cm

Time evolution $U(t)$ evolves all observables $\cb\equiv\cb^{(0)}$ 
continuously in 
conventional quantum physics:$\omega(U(t)^{-1}bU(t))$ is continuous in $t$ for
all $b\in\cb^{(0)}$.

Let $\cb^{(1)}\subset \cb^{(0)}$ be the subset of $\cb^{(0)}$ with 
differentiable time
evolution.The Hamiltonian $H$ is defined only on $\cb^{(1)}$: If $b(t)=
U(t)^{-1}b(0)U(t)\in \cb^{(1)} $ , then $idb(t)/dt$ $=[b(t),H]$.
For example, for $H=p^{2}/2m$ plus a smooth potential $V(x)$,$\cb^{(1)}$
contains twice-differentiable functions of $x$. For $D=-i\alpha.\partial$,
it has $C^{1}$ functions.

K-times differentiabilty in this way gives $\cb^{(K)}$ with inclusions $...
\subset \cb^{(K)}\subset\cb^{(K-1)}\subset...\subset\cb^{(0)}$.

Let $\cb^{(\infty)}$=$\bigcap\cb^{(K)}$.From $\cb^{(\infty)}$,we have to 
extract a 
subalgebra which helps us reconstruct the spatial slice $Q$ with its 
differential
structure,dimension etc. The criterion to do so may be a weak form of 
relativistic causality. In relativity,if an observable is localised in a 
spatial region $D$ at time zero, its support $D_{t}$ at time $t$ is within the 
future
light cone of $D$.This means in particular that as $t\rightarrow 0$,
$D_{t} \rightarrow D_{0}=D$.There is no spread all over in infinitesimal times.
Such a constraint is compatible with $H$ having a finite number of spatial
derivatives. Relativistic causality for example is violated by the Hamiltonian 
$(p^{2}+m^{2})^{1/2}$ whereas the Dirac operator is of first order and causal.

If $H$ is of first order and $f$ and $g$ are functions,then $[[H,f],g]=0$.
This is so for example for the Dirac Hamiltonian. More generally,if $H$
is of finite order, $[[H,f_1],f_2],f_3]...,f_K]]=0$ for a finite $K$.All
this suggests the

{\it Definition: A commutative subalgebra $\cc^{(\infty)}$ of 
$\cb^{(\infty)}$ is 
weakly causal if, for $f_i\in \cc^{(\infty)}$,
$[[H,f_1],f_2],f_3]...,f_K]]=0$ for some K.}

This pale form of causality can be valid {\it generically} only for functions
on a spatial manifold $M$. For example, the Hamiltonian of a simple harmonic 
oscillator fulfills this criterion in both position and momentum space.

{\em Conjecture:~$\cc^{(\infty)}$ determines $M$ and its $C^{\infty}$-structure 
by the analogue of a Gel'fand-Neumark construction.}

If $\cb^{(K)}$ is substituted for $\cb^{(\infty)}$ and a corresponding 
$\cc^{(K)}$
is extracted, the latter will fix only the $C^{K}$-structure of $Q$. 
Requiring just continuity , we can recover $Q$ only as a topological space.

We can expand observables in eigenstates of 
$H : b(t)=\sum b_{n}e^{i\omega_{n}t}$,
with $\parallel b(t)\parallel^2\equiv \omega(b(t)^*b(t))<\infty$.From 
$d^K b(t)/dt^K=\sum (i\omega)^{K}b_{n}e^{i\omega_{n}t}$, we see that 
requiring convergence of r.h.s. in  norm for high $K$ suppresses high
frequencies.(We are ignoring issues of null states of $\omega$ here.)Thus low
energy observations recover $Q$ with its $C^{\infty}$- structure. But as 
higher and higher energies are observed, that is, as shorter and shorter
time scales are resolved,$Q$ gets more rough, retaining progressively less of 
its differentable structure. Eventually for nondifferentiable $b$, $Q$ is
just a topological space and retains no differentiable structure.

The situation is in fact more dramatic.The algebra giving $Q$ as a topological
space is the $C^*$-algebra of continuous functions $\cc^{(0)}$.The maximum
commutative $C^*$-algebra $\cc^{(0)''}$ containing $\cc^{(0)}$ does not give
$Q$ as a topological space modelled on a manifold.

Much of what we discussed above is based on spectral considerations, 
suggesting that more remarks are necessary as regards isospectral manifolds.
We will not however undertake this task here.

\vglue 0.6cm
\sxn{ {\bf  Dimension and Metric}}
\vglue 0.5cm

Suppose that  $Q$ has been recovered as a 
manifold.  We can then find its dimension in the usual way.

There is also a novel manner to find its dimension $d$ from the spectrum
$\{{\lambda}_n\}$
of $H$: If $H$ is of order $N$, $|{\lambda}_n|$ 
grows like $n^{N/d}$ as $n \rightarrow \infty$ \cite{co,col1,con,fg}.

We can find a metric as well for  $Q$ \cite{co,con,fg}:  It is
specified by
the
distance 

\be
d(x,y)=\{\sup_{a}|a_c(x)-a_c(y)|:\frac {1}{N!} ||\underbrace {[a,[a,\ldots
[a,H]
\ldots ]}_{N ~a's}||\leq 1\}.
\ee

\noindent This remarkable formula gives the usual metric for the Dirac
operator
$[N=1]$ \cite{co,col1} and  the Laplacian $[N=2]$ \cite{fg}.

\vglue 0.6cm
\sxn{{\bf What is Quantum Topology?}}
\vglue 0.5cm

A question of the following sort often suggests itself when encountering
discussions of topology in quantum gravity:  If $Q$ is a topological
space, possibly with additional differential and geometric structures
[``classical'' data], what is meant by {\em quantizing} $Q$?

It is perhaps best understood as: {\em finding an algebra of operators on
a Hilbert space from which $Q$ and its attributes can be reconstructed}
[much as in the Gel'fand-Naimark theorem].

\vglue 0.6cm
\sxn{{\bf Topology Change}}
\vglue 0.5cm

We now use the preceding ideas to discuss topology change, following
ref. 3. [See ref. \cite{ma2} for related work.]

There are indications from theoretical considerations that spatial
topology in quantum gravity cannot be a time-invariant attribute, and
that its transmutations must be permitted in any eventual theory.

The best evidence for the necessity of topology change comes from the
examination of the spin-statistics connection for the so-called geons
\cite{fs,sor,bdgmss1,paulo,bmss}.  Geons are solitonic excitations caused by 
twists in spatial
topology.  In the absence of topology change, a geon can neither
annihilate nor be pair produced with a partner geon, so that no geon has
an associated antigeon.

Now spin-statistics theorems generally emerge in theories admitting
creation-annihilation processes \cite{sor,bdgmss1,ds}.  It can therefore be
expected to
fail for geons in gravity theories with no topology change.  Calculations
on geon quantization in fact confirm this expectation \cite{sor,abbjrs}.

The absence of a universal spin-statistics connection in these gravity
theories is much like its absence for a conventional nonrelativistic
quantum particle which too cannot be pair produced or annihilated.   Such
a particle can obey any sort of statistics including parastatistics
regardless of its intrinsic spin.  But the standard spin-statistics
connection can be enforced in nonrelativistic dynamics
also by
introducing
suitable creation-annihilation processes \cite{see}.

There is now a general opinion that the spin-statistics theorem should
extend to gravity as well.  Just as this theorem emerges from even
nonrelativistic physics once it admits pair production and annihilation
\cite{bdgmss1}, quantum gravity too can be expected to become compatible
with this
theorem after it allows suitable topology change \cite{ds}.  In this
manner,
 the desire for the
usual spin-statistics connection leads us to look for a quantum gravity
with transmuting topology.

Canonical quantum gravity in its elementary form is predicated on the
hypothesis that spacetime topology is of the form $\Sigma \times {\bf
{R}}$
(with
${\bf {R}}$ accounting for time) and has an eternal spatial topology.
This
fact has led to numerous
suggestions that conventional canonical gravity is inadequate if not
wrong, and must be circumvented by radical revisions of spacetime
concepts \cite{blms}, or by improved approaches based either on functional
integrals and cobordism \cite{ds} or on alternative quantization methods.

Ideas on topology change were first articulated in quantum gravity, and
more specifically in attempts at semiclassical quantization of classical
gravity.  Also it is an attribute intimately linked to gravity in the
physicist's mind.  These connections and the apparently revolutionary
nature of topology change as an idea have led to extravagant speculations  
about twinkling topology in quantum gravity and their impact
on fundamental concepts in physics.

Here we show that models of quantum particles exist which admit topology
change or contain states with no well-defined classical topology.  {\em
This is so even  though gravity does not have a central role in our ideas
and is significant only  to  the extent that metric is important for a
matter Hamiltonian.}
These models use only known physical principles and have no
revolutionary content, and at least suggest that topology change in
quantum gravity too may be achieved with a modest physical input and no
drastic alteration of basic laws.

We consider particle dynamics.
The configuration space of a particle being ordinary space, we are thus
imagining a physicist probing spatial topology using a particle.

Let us consider a particle with no internal degrees of freedom living on
the union $Q^{\prime}$ of two intervals which are numbered as 1 and 2:

\be
Q^{\prime}=[0,2\pi] \bigcup [0,2\pi] \equiv Q^{\prime}_1 \bigcup
Q^{\prime}_2~.\label{11.1}
\ee

\noindent It is convenient to write its wave function $\psi$ as $(\psi_1,
\psi_2)$,
where each $\psi_i$ is a function on $[0,2\pi]$ and $\psi^*_i \psi_i$ is
the probability density on $Q^{\prime}_i$. The scalar product between
$\psi$ and another wave function $\chi=(\chi_1,\chi_2)$ is

\be
(\psi,\chi)=\int_0^{2 \pi} dx \sum_i (\psi^*_i\chi_i)(x)~.
\label{11.2}
\ee

It is interesting that we can also think of this particle as moving on
$[0,2\pi]$ and having an internal degree of freedom associated with the
index $i$.

After a convenient choice of units, we define the Hamiltonian formally by

\be
(H \psi)_i(x)=-\frac{d^2 \psi_i}{dx^2}(x)~ \label{11.3}
\ee

\noindent [where $\psi_i$ is assumed to be suitably differentiable in the
interval
$[0,2\pi]$]. This definition is only formal as we must also specify its
domain
$\ch^1$
\cite{rs2}. The latter involves the statement of the boundary conditions
(BC's) at $x=0$ and $x=2\pi$.

Arbitrary BC's are not suitable to specify a domain: A symmetric operator
${\cal O}$
with domain $D({\cal O})$ will not be self-adjoint unless the following
criterion is also fulfilled:

\be
{\cal B}_{\cal O}(\psi,\chi)\equiv (\psi,{\cal O} \chi)-(
{\cal O}^{\dagger}\psi,\chi)=0~~{\it for~all}~\chi \in D({\cal O})
 \Leftrightarrow
\psi \in D({\cal {O}})~.
\label{11.4}
\ee

For the differential operator $H$, the form ${\cal B}_{H}( \cdot, \cdot)$
is
given by

\be
{\cal B}_H(\psi,\chi)=
\sum_{i=1}^{2} \left [ -\psi^{*}_{i}(x) 
\frac{d \chi_{i}(x)}{dx}+
\frac{d\psi^{*}_{i}(x)}{dx}\chi_{i}(x)\right ]^{2 \pi}_{0}~.\label{11.5}
\ee

\noindent It is not difficult to show that there is a $U(4)$ worth of
$D(H)\equiv\ch^1$ here
compatible with (\ref{11.4}).

We would like to restrict this enormous choice for $D(H)$, our intention
not being to study all possible domains for $D(H)$. So let us 
restrict ourselves to the domains

\be
D_{u}=\{\psi \in C^2(Q^{\prime}):\psi_i(2 \pi) =
u_{ij}\psi_j(0),~~
\frac{d\psi_{i}}{dx}(2\pi)=u_{ij}
\frac{d\psi_{j}}{dx}(0),~u \in U(2)\}~. \label{11.6}
\ee

\noindent  These domains have the
virtue
of being compatible with the definition of momentum in the sense discussed
in ref. 2.

There are two choices of $u$ which are of particular interest:

\be
a)~~~~~~~~~~
 u_{a}=\left( \begin{array}{cc}
0 & e^{i\theta_{12}} \\
e^{i\theta_{21}} & 0
\end{array}\right) , \label{11.7}
\ee

\be
b)~~~~~~~~~
 u_{b}=\left( \begin{array}{cc}
e^{i\theta_{11}} & 0 \\
0 & e^{i\theta_{22}}
\end{array} \right) .\label{11.8}
\ee

In case $a$, the density functions $\psi^*_i \chi_i$ fulfill
\be
(\psi^*_1 \chi_1)(2\pi)=(\psi^*_2 \chi_2)(0)~,
\ee

\be
(\psi^*_2 \chi_2)(2 \pi)=(\psi^*_1 \chi_1)(0)~.\label{11.9}
\ee

\noindent Figure 1 displays (\ref{11.9}), these densities being the same at the
points
connected by broken lines.

  \begin{figure}[hbtp]
  \epsfysize=4cm
  \epsfxsize=6cm
 \centerline{\epsfbox{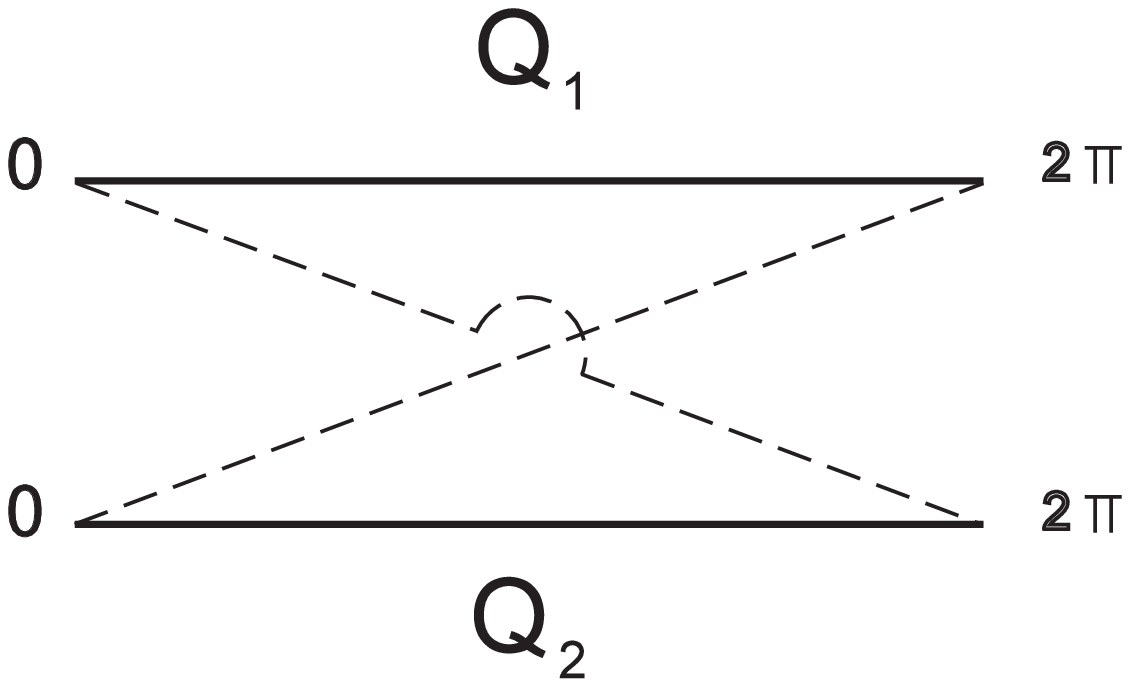}}
 \caption{  In case $a$, the density
functions are the same at the points joined by broken lines in this Figure.}
\end{figure}

In case $b$, they fulfill, instead,
\be
(\psi^*_1 \chi_1)(2\pi)=(\psi^*_1 \chi_1)(0)~,
\ee

\be
(\psi^*_2 \chi_2)(2 \pi)=(\psi^*_2 \chi_2)(0)~\label{11.10}
\ee

\noindent which fact is shown in a similar way in Figure 2.

\begin{figure}[hbtp]
  \epsfysize=4cm
  \epsfxsize=6cm
 \centerline{\epsfbox{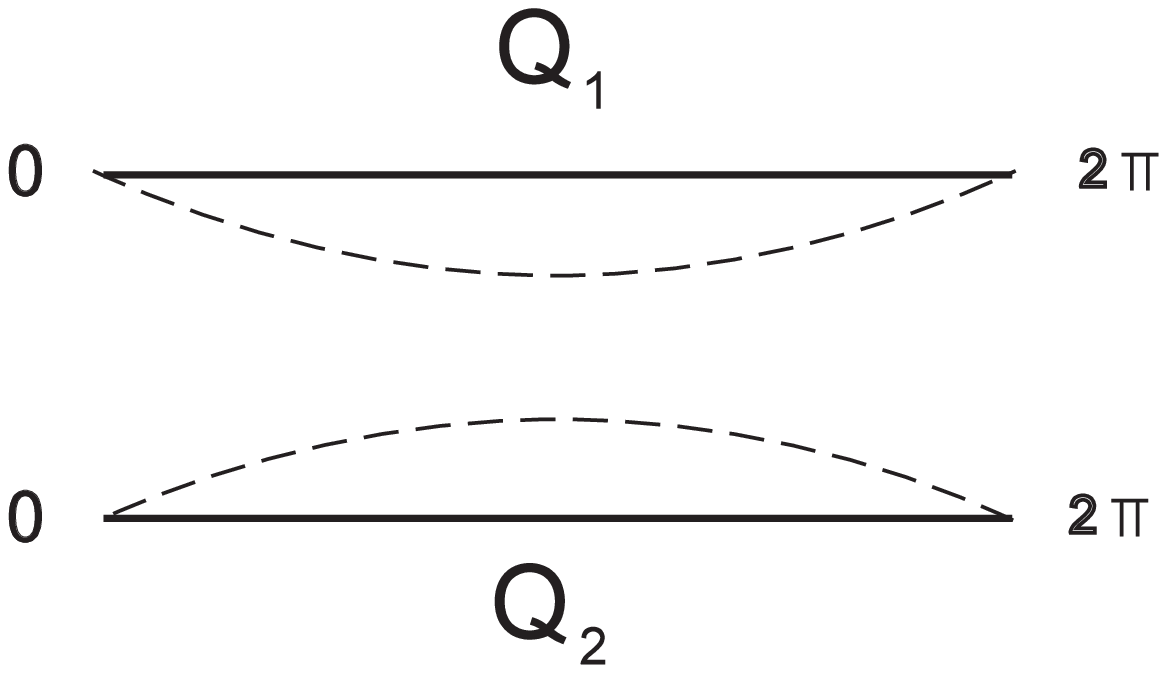}}
 \caption{ In case $b$, the density
functions are the same at the points joined by broken lines in this Figure.}
\end{figure}

Now if  $\psi^*_i ,\c_i\in D_u$,then $\psi^*_i \chi_i\in \cc^{(0)}$ in the 
operator-theoretic approach used earlier.Such probabilty densities 
in fact generate $\cc^{(0)}$.Therefore their continuity properties determine 
the
topology of the 
space to be identified as $Q$.It follows that we can identify the
points joined by dots to get the classical configuration space $Q$
if $u=u_a$ or $u_b$. It is
{\it not} $Q^{\prime}$, but rather a circle $S^1$ in case $a$ and
the union $S^1 \bigcup S^1$ of two circles in case $b$.

The requirement $H^M D_u^{\infty} \subseteq D_u^{\infty}\subset D_u $ for 
$u=u_{a,b}$ and for all 
$M\in {\IN}$ implies that
arbitrary derivatives of $\psi^*_i \chi_i\in D_u^{\infty} $ are continuous 
at the points
joined by broken lines, that is on $S^1$ and $S^1 \bigcup S^1$ for the two
cases.
We can prove this easily using (\ref{11.6}).
In this way, from $D_u^{\infty}$, we also recover $S^1$ and $S^1\bigcup S^1$ 
as manifolds.

When $u$ has neither of the values (\ref{11.7})  and (\ref{11.8}), then $Q$
becomes the union of two intervals. The latter
happens for example for

\be
u=\frac{1}{\sqrt 2}\left( \begin{array}{cc}
1 & 1 \\
-1 & 1
\end{array}\right)~~. \label{11.11}
\ee

\noindent In all such cases, $Q$ can be regarded as a manifold with
boundaries as shown
by the argument above.

\vglue 0.6cm
\noindent {\em \bf  Dynamics for Boundary Conditions}
\vglue 0.5cm

We saw in the previous section that topology change can be achieved in
quantum physics by treating the parameters in the BC's as suitable
external parameters which can be varied. 

However it is not quite satisfactory to have to regard $u$ as an external
parameter and not subject it to quantum rules.  We now therefore promote
it to an operator, introduce its conjugate variables and modify the
Hamiltonian as well to account for its dynamics.  The result is a closed
quantum system.  It has no state with a sharply defined $u$.  We cannot
therefore associate one or two circles with the quantum particle and
quantum spatial topology has to be regarded as a superposition of
classical spatial topologies.  Depending on our choice of the
Hamiltonian, it is possible to prepare states where topology is peaked at
one or two $S{^1}$'s for a long time, or arrange matters so that there is
transmutation from one of these states to another.

Quantization of $u$ is achieved as follows. Let $T(\alpha)$ be the
antihermitean generators of the Lie algebra of $U(2)$ [the latter being
regarded as the group of
$2
\times 2$ unitary matrices] and normalized according to $Tr~ T(\alpha)
T(\beta)=-N \delta_{\alpha \beta}$, $N$ being a constant. Let $\hat u$ be the
matrix of quantum operators representing the classical $u$. It fulfills

\be
\hat u_{ij} \hat u^{\dagger}_{ik} = {\bf 1} \delta_{jk},~~[\hat u_{ij}, \hat
u_{kh}]=0~, \label{11.15}
\ee
$\hat u_{ik}^{\dagger}$ being the adjoint of $\hat u_{ik}$.
The  operators conjugate to $\hat u$ will be denoted by $L_{\alpha}$. If
\be
[T_{\alpha},T_{\beta}]=c_{\alpha \beta}^{\gamma}T_{\gamma},
\ee

\be
c_{\alpha \beta}^{\gamma}={\rm structure~constants~of}~U(2),
\label{11.16}
\ee
$L_{\alpha}$ has the commutators
\be
[L_{\alpha},\hat u]=-T(\alpha)\hat u~, 
\label{11.16.5}
\ee

\be
[L_{\alpha},L_{\beta}]=c_{\alpha\beta}^{\gamma}L_{\gamma}, \label{11.17}
\ee

\be
[T(\alpha)\hat u]_{ij}\equiv T(\alpha)_{ik}\hat u_{kj}.
\ee

If $\hat V$ is the quantum operator for a function $V$ of $u$,
$[L_{\alpha},\hat V]$ is
determined by (\ref{11.16.5}) and (\ref{11.17}). 

The Hamiltonian for the combined particle-$u$ system can be taken to be,
for example,
\be
\hat H = H + \frac{1}{2I}\sum_{\alpha} L^2_{\alpha},\label{11.18}
\ee

\noindent $I$ being the moment of inertia.

Quantized BC's with a particular dynamics are described by (\ref{11.15}),
(\ref{11.16.5}),(\ref{11.17}) and (\ref{11.18}).

The general state vector in the domain of $\hat H$ is a superposition of
state vectors $\phi \otimes_{\bf C}|u\rangle $ where $\phi \in D_u$ and 
$|u\rangle$ is a generalized eigenstate of $\hat u$:

\be
\hat u_{ij}|u\rangle=u_{ij}|u\rangle,~~\langle u^{\prime}|u
\rangle=\delta(u^{\prime -1}u)~.\label{11.20}
\ee
The $\delta$-function here is defined by

\be
\int du f(u)\delta(u^{\prime -1}u)=f(u^{\prime}),\label{11.21}
\ee

\noindent $du$ being the (conveniently normalized) Haar measure on $U(2)$.

It follows that the classical topology of one and two circles is
recovered on the states $\sum_\lambda  C_{\lambda}
\phi_{u_a}^{(\lambda)}\otimes_{\bf C}|u_{a}\rangle $ and
$\sum_\lambda
D_{\lambda}\phi_{u_b}^{(\lambda)}\otimes_{\bf
C}|u_b\rangle,~[C_{\lambda},~D_{\lambda}\in
{\comm}, ~\phi_{u_{a,b}}^{(\lambda)}\in D_{u_{a,b}}$] with the two fixed
values $u_a$,  and $u_b$ of (\ref{11.7}) and (\ref{11.8}) for $u$.

As the dynamical system has been enhanced by $U(2)$, the configuration
space we recover is not $Q$ in the strict sense,but rather $Q\times U(2)$.
But we will refer to only $Q$ as the configuration space below as a matter of 
convenience.

Now the above vectors are clearly idealized and unphysical , and with 
infinite norm.
The best we can do with normalizable vectors to localize topology around
one or two circles is to work with the wave packets

\be
\int du f(u) \phi_{u} \otimes_{\bf C} |u\rangle~,
\ee

\be
\int \phi_{u}\in D_u,
\ee

\be
\int du |f(u)|^2 < \infty \label{11.23}
\ee

\noindent where $f$ is sharply peaked at the $u$ for the desired topology.
The
classical topology recovered from these states will only approximately be
one or two circles, the quantum topology also containing admixtures from
neighboring topologies of  two intervals.

A localized state vector of the form (\ref{11.23}) is not as a rule an
eigenstate of a Hamiltonian like $\hat H$. Rather it will spread in
course of time so that classical topology is likely to
 disintegrate
mostly into that of two intervals.  We can of
course localize it around one or two $S^1$'s for a very long time by
choosing $I$ to be large, the classical limit for topology being achieved
by letting $I \rightarrow \infty$. By adding suitable potential terms, we
can also no doubt arrange matters so that a wave packet concentrated
around $u=u_a$ moves in time to one concentrated around  $u=u_b$. This
process would be thought of as topology change by a classical physicist.

The preceding considerations on topology change admit generalizations to
higher dimensions as explained in ref.3.

\vglue 0.6cm
\sxn{ {\bf  Final Remarks}}
\vglue 0.5cm

In this article we have touched upon several issues concerning quantum
topology and showed their utility for research of current interest such
as topology change and fuzzy topology.  Our significant contribution, if
any, here has been in formulating new fundamental problems with
reasonable clarity.  We have also sketched a few answers, but they are
tentative and incomplete.

\vglue 0.6cm
\sxn{{\bf Acknowledgments}}
\vglue 0.5cm

The work reported in this article is part of an ongoing program with
several colleagues.  I have especially benefited from discussions with
Jan Ambjorn,Peppe Bimonte, T.R. Govindarajan, Gianni Landi, Fedele Lizzi,
Beppe Marmo,
Shasanka Mohan Roy, Alberto Simoni and Paulo Teotonio-Sobrinho in its
preparation.I am also deeply grateful to Arshad Momen for his 
extensive and generous help in the preparation of this paper.  This work 
was supported by the U.S. Department of Energy
under Contract Number DE-FG-02-85ER40231.

 \end{document}